# Title: Quantum-inspired super-resolution of fluorescent point-like sources

**Authors:** Cheyenne S. Mitchell[1,2], Dhananjay Dhruva[2,3,4], Zachary P. Burke[1], David J. Durden[1,2], Armine I. Dingilian[2,3,4], Mikael P. Backlund[1,2,3,4*]

**Affiliations:**

[1]Department of Chemistry, University of Illinois at Urbana-Champaign; Urbana, IL 61801, USA.

[2]Illinois Quantum Information Science and Technology Center (IQUIST), University of Illinois at Urbana-Champaign; Urbana, IL 61801, USA.

[3]Center for Biophysics and Quantitative Biology, University of Illinois at Urbana-Champaign; Urbana, IL 61801, USA.

[4] NSF Science and Technology Center for Quantitative Cell Biology, University of Illinois at Urbana-Champaign; Urbana, IL 61801, USA.

*Corresponding author. Email: mikaelb@illinois.edu

**Abstract:** We report the experimental super-resolution of pairs of point-like fluorescent sources using a modified image inversion interferometer microscope. The technique is inspired by recent developments in the application of quantum parameter estimation theory to semiclassical imaging problems. We find that the image inversion technique requires special polarization filtering to account for the dipolar nature of the emission. Using an azimuthal polarizer, we obtain improvements in the Fisher information of source separation by over an order of magnitude relative to direct imaging. Unlike established super-resolution fluorescence techniques, the method does not require sequential photoswitching/blinking of the fluorophores, and thus could facilitate significant speed-ups for certain biological imaging/tracking tasks.

**Main Text:**

In the context of optical imaging, resolving objects separated by distances less than the wavelength of light is fundamentally difficult (Fig. 1A-B). This basic fact of physics has been appreciated since it was articulated in the works of Abbe (*1*) and Rayleigh (*2*) during the latter half of the nineteenth century. According to Rayleigh's criterion, when the separation of two point sources is less than ~0.61 $\lambda/$ NA (where $\lambda$ is the wavelength of light and NA is the imaging system's numerical aperture) they are rendered unresolvable. The development of classical statistical inference in the century-plus since Rayleigh calls for some refinement of this statement, since given enough measurement averaging (i.e. a sufficient number of detected photons) one can always discern between the binary hypotheses of one vs. two emitters with an arbitrarily small error rate. One popular reformulation of Rayleigh's criterion invokes the Fisher information (FI) (*3*), a figure-of-merit that governs one's ability to extract a precise estimate of one or more parameters from a set of noisy observations (*4*). "Rayleigh's curse" says that the FI with respect to estimating the separation distance between two emitters vanishes as the separation goes to zero (*5*, *6*). Equivalently, the inverse of the FI, the Cramér-Rao bound (CRB), diverges in the same limit.

In the broader context of quantum parameter estimation theory, the classical FI is bounded above by the more fundamental quantum Fisher information (QFI) (*7*). From this vantage, Tsang, Nair, and Lu recently proved that Rayleigh's curse is illusory in some sense (*5*). Whereas the FI associated with "direct imaging" (i.e. a pixel-resolved intensity measurement at a conjugate image plane) vanishes in the limit of zero separation, the QFI in fact remains constant. This means that some measurement scheme other than direct imaging can in principle resolve two point objects deep in the subdiffraction regime much more efficiently than can be done via direct imaging. Said differently, direct imaging effectively leaves unharvested information contained in the phase of the light. An appropriately chosen phase-sensitive measurement can recover this lost information. To be clear, even though the calculation of QFI begins with the density operator describing the quantum electromagnetic field, overcoming Rayleigh's curse does not depend on multi-particle entanglement. It only requires phase coherence in the light. Indeed, the result can be retrofitted into the language of semiclassical detection theory where the hidden information is contained in the phase of the classical electromagnetic field (*8*, *9*). Tsang and coworkers' seminal paper has since spawned a flurry of follow-up theoretical and experimental work as reviewed in Ref. (*6*); the emerging area is collectively referred to as "quantum-inspired super-resolution".

From the outset, Tsang, Nair, and Lu proposed a measurement based on spatial-mode demultiplexing (SPADE) to saturate the QFI (*5*, *6*, *9*–*15*). For a Gaussian point-spread function (PSF), this corresponds to photon counting in the basis of Hermite-Gaussian transverse modes (*5*). For a more general PSF (e.g., an Airy disk), the optimal basis for SPADE can be constructed algorithmically (*16*, *17*). A related proposal based on image inversion interferometry (which we will refer to as "III" but has also been called SLIVER) effectively sorts into even- and odd-parity transverse spatial modes, which is in theory sufficient to break Rayleigh's curse for two closely spaced point emitters (*14*, *18*, *19*). The idea is that a single point source located on the optical axis will produce an even-parity field, and as such all the collected light will be shunted to one of the two interferometer outputs. Now if the position deviates slightly from the optical axis, a few photons will be recorded in the odd-parity channel, such that by sitting on this dark fringe the measurement can be made very sensitive to displacement of the source from the optical axis. The last ingredient in translating this into source-pair super-resolution is to align the system such that the optical axis coincides with the average position of the pair (which is considerably easier to estimate from direct imaging than is separation). While SPADE and its variants promise a new route to super-resolution imaging in microscopy, telemetry, and everything between, the experimental demonstrations published to date have tended to target a pair of laser spots with spoiled mutual coherence (*12*, *13*, *18*, *20*–*22*). In this work we report the experimental demonstration of quantum-inspired super-resolution in the context of fluorescence microscopy.

Of course, a suite of Nobel-prize-winning super-resolution fluorescence microscopy techniques already exist in the forms of PALM/STORM (*23*–*25*), STED (*26*, *27*), and variations thereof (*28*, *29*). Before proceeding we must place quantum-inspired super-resolution against this backdrop. Both the PALM/STORM and STED routes to super-resolution depend on the controlled photoswitching of fluors. For PALM/STORM in particular this means stochastically separating the emission events of neighboring molecules in time, and as a result these methods are often prohibitively slow. MINFLUX microscopy has recently been used to achieve unprecedented resolutions (*30*–*33*), but still typically relies on stochastic switching of the labels. By contrast, quantum-inspired schemes do not require photoswitching, and so can in principle be used to super-resolve much more dynamic scenes labeled using a broader class of fluorescent emitters.

## Concept and Theory

Figure 1C in part restates Tsang, Nair, and Lu's main theoretical result (*5*), but with one important revision. The original theory invoked the scalar approximation to effectively treat a pair of monopole emitters. In nature, however, there are no monopole emitters of electromagnetic radiation. A real "isotropic" point source consists of an ensemble of randomly oriented dipole emitters, the emission of which can be modeled as a mixture of *x*-, *y*-, and *z*-oriented dipole emitters [See Materials and Methods as well as Refs. (*34*–*37*)]. The data depicted in Fig. 1C result from a full vectorial diffraction calculation, wherein both isotropic point sources are modeled as a mixture of dipole emitters. For direct imaging, the result is qualitatively similar. The FI for direct imaging (Fig. 1C, blue line) vanishes at small separation despite the constant QFI. At a separation of 5 nm, the QFI is more than 4 orders of magnitude larger than the FI of direct imaging, meaning that there exists some measurement scheme other than direct imaging that can achieve a given resolution benchmark with fewer than $10^{-4}$ times as many detected photons. For our experimental pursuit of quantum-inspired super-resolution, we elected to construct a microscope based on image inversion interferometry (III, Fig. 2). However, a calculation of the FI of separation based on III as previously proposed yielded some initially sobering results (Fig. 1C, gold line). Whereas the proposed III would yield significant enhancement in the resolution of monopole emitters, it evidently only leads to a modest improvement over direct imaging when targeting more realistic isotropic emitters. To wit, at 5 nm separation the FI associated with III is only about a factor of 1.6 larger than that of direct imaging. The loss of performance can be understood in terms of the symmetry of the fields emitted by dipoles of different orientations (Figs. S1-S3). Taking *z* to be parallel to the optical axis (Fig. S1), emission from a dipole oriented perpendicular to *z* produces emission that is symmetric with respect to inversion about the centroid (Fig. S2 A-B, E-F). Thus a dipole oriented perpendicular to *z* and positioned on the optical axis will produce a null in one output port of the III (Fig. S3 A-B, E-F), just the same as it would for a monopole emitter. On the other hand, a dipole oriented parallel to *z* emits a field that is anti-symmetric with respect to inversion about the centroid (Fig. S2 C, G). A dipole parallel to *z* and positioned on the optical axis will produce a null in the *other output port* of the III (Fig. S3 C, G). If you know your dipole sources are oriented either perpendicular or parallel to *z*, then the III would in principle work as advertised. However, if the collected light comes from the emission of a dipole with some intermediate orientation or a mixture of dipoles with different-symmetry emission, then neither output port can be nulled perfectly (Fig. S2 D, H and Fig. S3 D, H). The resulting inability to "sit on a dark fringe" is precisely why the original design of the III fails to produce much improvement over direct imaging for a real isotropic emitter. We comment here that this effect should only be important for high-NA imaging systems. When the NA is very low, the fraction of light collected from dipoles bearing a significant component parallel to *z* is small, and so in this limit the collected field is effectively inversion symmetric. However, state-of-the-art fluorescence microscopes almost always employ high-NA objectives, meaning the effect of incomplete nulling cannot be ignored in this case. The calculations depicted in Fig. 1B model the NA = 1.45 oil objective we use in our experiments.

Thankfully, we soon recognized a way to salvage the super-resolving power of the III microscope by appropriate filtering of the polarization. At the Fourier plane of the microscope, if one resolves the light collected due to dipolar emission into the azimuthal and radial polarization basis, it is known that the azimuthally polarized portion is guaranteed to be anti-symmetric with respect to inversion, and that all of the asymmetry, if there is any, will be carried entirely by the radially polarized light (*38*, *39*). Thus, by throwing away the radially polarized light, one can recover the ability to sit on a dark fringe. Filtering the radially polarized light emitted by an isotropic source in this way and then injecting into the III results in an FI for source separation that is vastly improved relative to both direct imaging and unpolarized III (Fig. 1C, red line). Only a modest factor of about 2.6 separates the FI of this polarized III scheme from the QFI at small separations. The small persisting gap can be closed by mining the scraps of information contained in the radially polarized emission, but for an isotropic point source this is almost certainly more trouble than it's worth.

**Results and Discussion**

To experimentally realize this polarization-filtered version of III we constructed the setup depicted in Fig. 2. Two Dove prisms oriented orthogonal to one another and placed in opposite arms of the interferometer effect the image inversion. To achieve the desired polarization filtering, we added a vortex half wave plate to the III at a particular plane in the collection path. The vortex plate converts azimuthally polarized light into $y$ and radial into $x$ (*39*). The asymmetric part of the light can then be rejected by inserting a linear polarizer. The vortex phase shift renders the image of a single point source as a donut, akin to the excitation PSF in MINFLUX (*30*–*33*) and the stimulated emission beam in STED (*26*, *27*).

The experimental procedure we employed essentially constitutes an analog simulation of a pair of mutually incoherent point-like isotropic emitters separated by a range of subdiffraction distances (see Figs. S4-S5 and Materials and Methods section). We recorded tens-of-thousands of images of isolated 40-nm fluorescent beads as their positions were scanned in the proximity of the optical axis using a piezo stage. Images of the same bead recorded at opposing positions were combined in post-processing to emulate a source pair. A "ground truth" of separation was established by first recording calibration images in which the image of a target bead is well separated from its inverted dual. The target bead and a separate fiducial bead were both individually super-localized by fitting to a simplified PSF model. Then the target was scanned near the optical axis such that the inverted image pair overlapped and interfered with one another. In these images, the still-isolated fiducial was super-localized, and from this localization we triangulated the "ground-truth" position of the target relative to the optical axis.

Recorded images were combined and analyzed using a custom MATLAB routine, the output of which is a look-up table (LUT) of filtered images of source pairs on a grid of subdiffraction separations. A subset of this LUT is depicted in Fig. 3A-E for a sampling of subdiffraction separations, $\Delta x$, oriented parallel to the horizontal axis of the image. (Additional samplings of the library can be found in Figs. S6-S7.) In Channel 1, constructive interference yields a single bright donut that does not change much as a function of small separations. The lion's share of the information is conveyed in Channel 2, where destructive interference leads to near-perfect nulling when the separation approaches zero. As the separation increases, a bowtie shape appears and becomes increasingly bright. The experimentally-derived LUT of processed images agrees well with simulated images of isotropic point source pairs (Fig. S8).

For fair comparison we repeated our experimental procedure with two important modifications. First, the vortex half wave plate and linear polarizer were removed such that the image of a point source again resembled a 2D Gaussian. Second, one of the delay stages within the interferometer was displaced such that the coherence between the two arms was destroyed. In this way we could implement direct imaging in a manner that lends itself to analysis with exactly the same analysis pipeline. For reference, the results of this control are shown in Fig. 3F-J. One can immediately appreciate that direct imaging will struggle in resolving source pairs by observing how little the images change across the panels.

To lend some additional quantitation to the assessment of our III microscope we computed the fringe visibility as defined by:

$$V\left(\Delta x,\Delta y\right)=\frac{\sum_k I_k^{(1)}\left(x_k,y_k;\Delta x,\Delta y\right)-\sum_k I_k^{(2)}\left(x_k,y_k;\Delta x,\Delta y\right)}{\sum_k I_k^{(1)}\left(x_k,y_k;\Delta x,\Delta y\right)+\sum_k I_k^{(2)}\left(x_k,y_k;\Delta x,\Delta y\right)} \quad (1)$$

Where $I_k^{(1)}\left(x_k,y_k;\Delta x,\Delta y\right)$ and $I_k^{(2)}\left(x_k,y_k;\Delta x,\Delta y\right)$ indicate the background-corrected intensity in the $k^{\text{th}}$ pixel of the images in Channels 1 and 2, respectively. In Eq. (1), $\left(\Delta x,\Delta y\right)$ denote the "ground-truth" separation estimated from triangulating the positions of the fiducial. Results are plotted in Fig. 3K. For comparison, the visibilities obtained from our direct imaging control are displayed in Fig. 3L. Additional visibility data are given in Fig. S9 as a function of delay stage position under different interferometer configurations.

Digging deeper into the data, our image libraries allow us to compute the realized FI with respect to separation for each imaging modality. Figure 4A sketches the experimentally recovered FIs for polarized III imaging (red) as well as for the direct imaging control (blue). Importantly, we note that the FI for polarized III in this display has already been diminished by a factor of 0.36 in order to assess a penalty for throwing away the radially polarized light. At a separation of 5 nm, the FI associated with polarized III is roughly 17x larger than that of the direct imaging control. We evidently do not realize the full 4 orders of magnitude improvement predicted by theory, but we nonetheless improve by more than one order of magnitude. In both cases we found that the FIs computed in this way were lower than those predicted from calculations (Fig. 4A, dashed lines), presumably due to experimental non-idealities that are not captured by the theoretical model, such as residual phase aberrations, pupil apodization, and excess noise from the camera electronics. The finite size and residual anisotropy of the fluorescent beads likely also contribute to this discrepancy.

As an additional control, we sought to parse the observed improvement into contributions from the change in PSF shape vs. that due to the interferometry. We repeated the experiment with the vortex half wave plate and linear polarizer in place, but with one of the delay lines in the interferometer displaced to spoil the coherence. The experimentally recovered FI for this "direct donut" imaging modality is about equal that of ordinary direct imaging at 5 nm separation (Fig. S10). We conclude that the resolution enhancement realized in comparing the polarized III to ordinary direct imaging can be attributed to the interferometer rather than the PSF shape.

We evaluated experimental performance in yet another way by generating separation estimates from carefully paired noisy images, then computing the mean-squared error (MSE) of this estimate relative to the "ground truth" obtained from triangulating the position of the fiducial. For ordinary direct imaging we estimated separation based on a least-squares fit to a pair of Gaussian spots. For the polarized III data, we compared the processed image library from one day's worth of data to the noisy images recorded on a different day, then produced an estimate of separation by finding the minimum squared error between the model and the noisy image. Results from alternative estimators and/or analysis pipelines are depicted in Fig. S11, showing similar behavior at small separations. At the particular signal-to-noise ratio we worked at (which was throttled by the need to collect many images over an extended period of time), we found that the separation estimator used for direct imaging exhibited significant biases. Since CRB (the inverse of the FI) is guaranteed only to lower-bound the variance of any *unbiased* estimator, we might not necessarily expect the MSEs depicted in Fig. 4B to look like the inverses of the curves in Fig. 4A. Nonetheless, at 5 nm separation we find that ordinary direct imaging produced a mean MSE about 16x worse than that derived from polarized III.

**Conclusions**

We have revised the theory of image inversion interferometry to account for the direction- and polarization-dependent emission of real point sources. With appropriate polarization filtering, the III can indeed beat Rayleigh's curse. We constructed an III microscope and carefully benchmarked its experimental performance, effectively demonstrating quantum-inspired super-resolution imaging. While the experimentally realized resolution enhancement is significant, theory predicts a great deal of room to improve. One upgrade to the setup to be implemented in the future is to correct phase aberrations individually in both arms of the interferometer, as suggested in (*40*). We also suspect that asymmetric apodization due to imperfections in one or more optical components may have limited our performance. The finite size of the fluorescent beads used (40 nm) likely diminished the apparent resolution of each imaging modality as well. Smaller beads or quantum emitters could be used, but at the cost of reduced brightness and/or photostability.

Our microscope enables a new route to super-resolution in fluorescence microscopy that does not depend on photoswitching of the emitters. This expands the list of compatible fluorescent labels and could lead to significant speed-ups in certain biological imaging/tracking applications. Of course, our experiments only demonstrate super-resolution on a very simple scene consisting of two point sources. It's unclear how its performance might translate to more complicated scenes (*17*, *41–44*). Thus, in cases where very little prior information is available about the scene, established super-resolution fluorescence techniques will likely still reign supreme. On the other hand, in cases where one knows their scene consists of just two sources (e.g., in tracking gene loci in diploid cells (*45*)), the method could prove powerful. We also expect to find utility in efficiently estimating the size, aspect ratio, and orientation of subdiffraction extended objects (*9*, *10*).

---

**Acknowledgements:** We thank Dongbeom Kim, Gina Lorenz, Paul Kwiat, Simeon Bogdanov, Ali Passian, Sam Bhagia, and Patrick Snyder for fruitful discussions.

**Funding:** This research was supported by the Department of Energy grant DE-SC0023167 (to CSM and MPB). Additional support due to the National Science Foundation grant 2243257 (to DD, AID, and MPB).

**Author contributions:** CSM, DD, and ZPB constructed the apparatus and performed the experiments. MPB, CSM, and DD analyzed the data. MPB performed the theory and computations. DJD provided support in interfacing software and hardware. AID provided theory support. MPB conceived of the project.

The authors declare no competing interests.

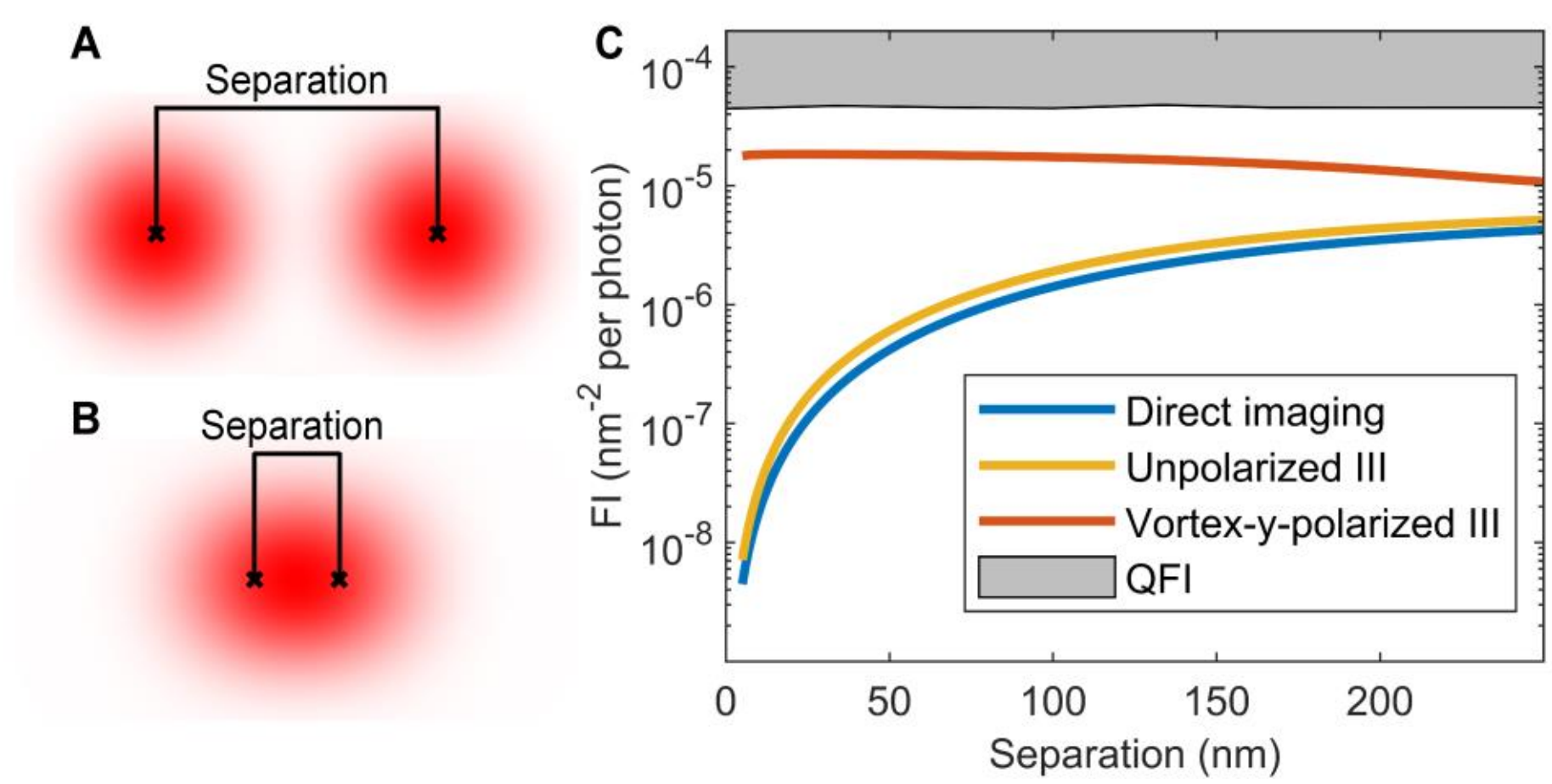

**Fig. 1: Illustration of point source separation and theoretical Fisher information calculations.** Calculated direct images of a pair of isotropic point emitters at (**A**) easily resolved and (**B**) subdiffraction separations. (**C**) Fisher information with respect to separation of a pair of realistic isotropic emitters, derived from vectorial diffraction calculations. The gray box is bounded below by the quantum limit. Our polarized image inversion scheme nearly saturates the quantum bound (red), while both unpolarized image inversion (gold) and direct imaging (blue) fair much worse.

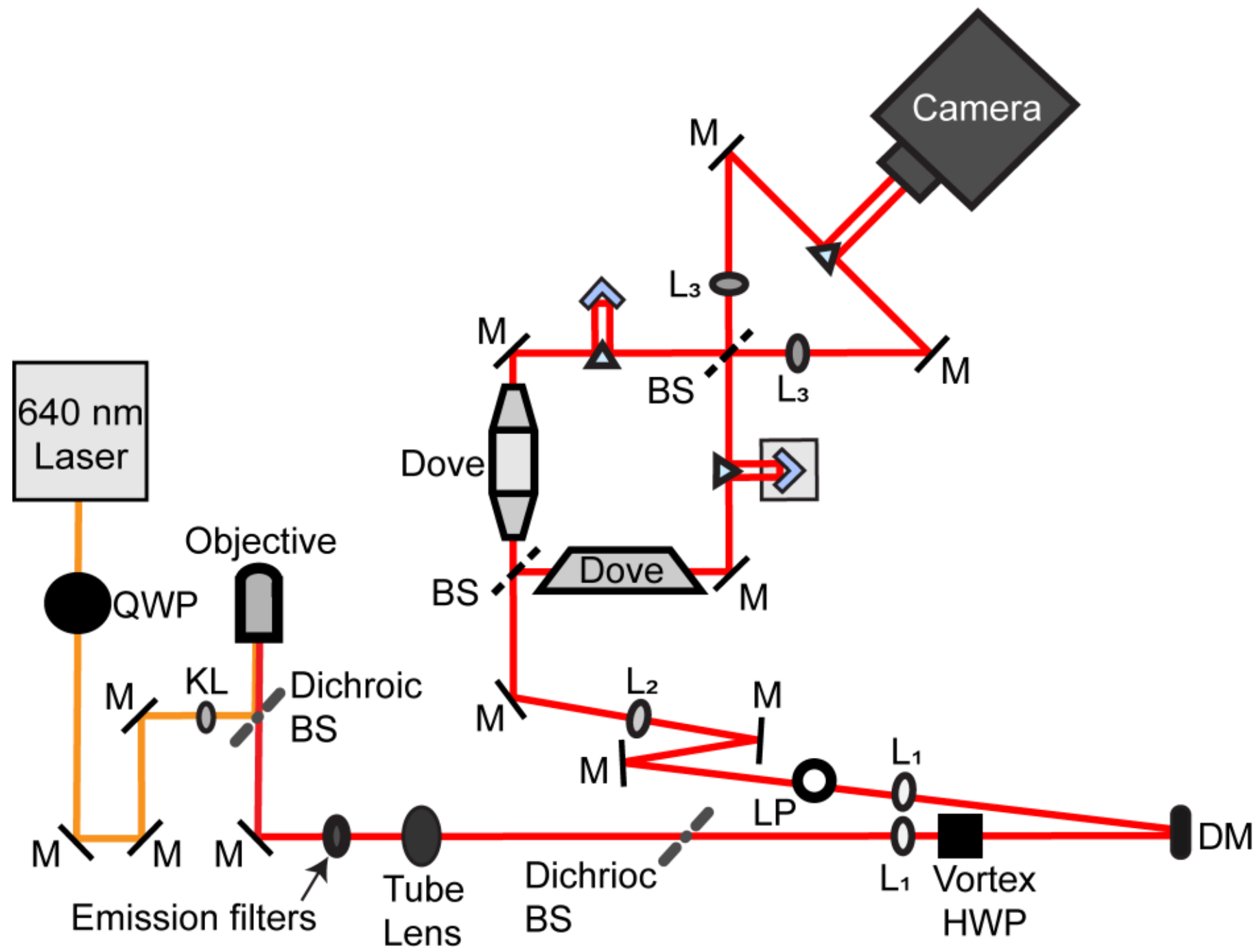

**Fig. 2: Experimental Setup**

Schematic of setup used to realize quantum-inspired super-resolution via polarized image inversion interferometry. QWP: quarter wave plate, M: mirror, KL: Köhler lens, BS: beam splitter, L: lens, Vortex HWP: vortex half wave plate, DM: deformable mirror, LP: linear polarizer. The second dichroic BS placed just after the tube lens compensates for birefringence due to the first dichroic BS. The deformable mirror is used to correct for phase aberrations. The vortex HWP and LP combine to reject the radially polarized emission.

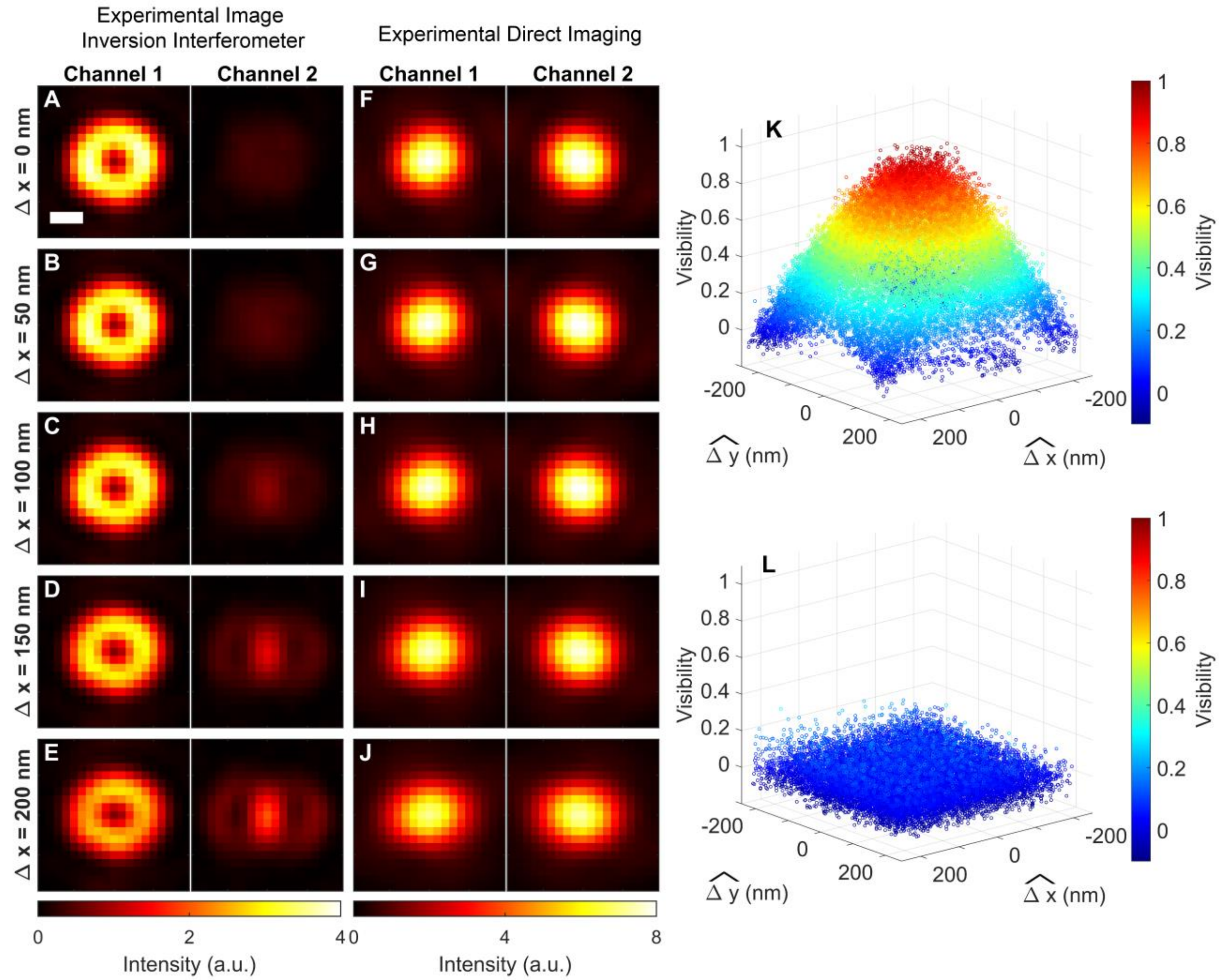

**Fig. 3: Experimentally derived images and associated fringe visibilities.** (**A-E**) Depicts the processed images resulting in both of the two channels of the polarized image inversion microscope for various point source separations. (**F-J**) Shows that corresponding processed images for a control experiment in which the polarizing elements were removed and the coherence between the arms of the interferometer was spoiled, effectively producing two direct imaging experiments, the information from which is recombined to produce the solid blue line in Fig. 4A. Note the different intensity colormaps for (A-E) vs. (F-J). (**K**) Experimental fringe visibility realized by the polarized image inversion microscope. (**L**) Fringe visibility of the control. Scale bar: 250 nm.

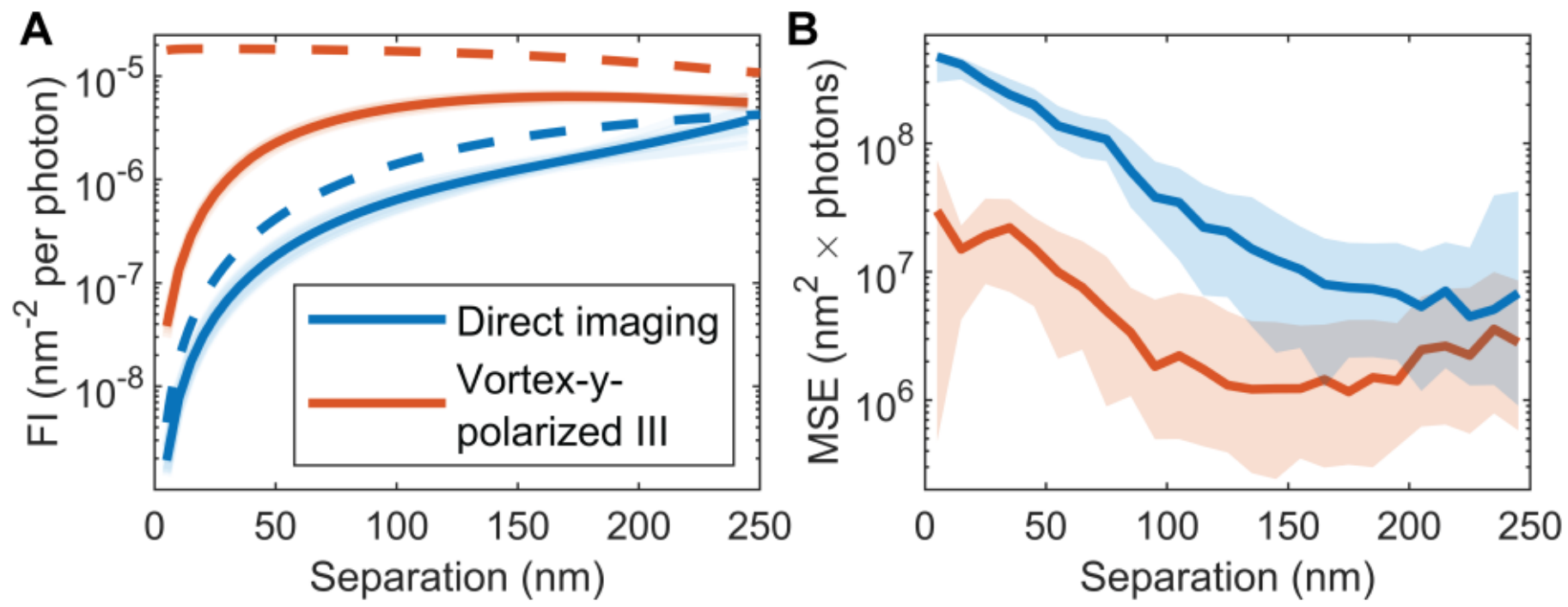

**Fig. 4: Experimental results.** (**A**) The realized Fisher information, as derived from processed experimental images of the polarized image inversion interferometer (red) and direct imaging control (blue). Transparent lines: FI computed for separation vectors oriented along each of 10 directions between 0° and 180°. Opaque lines: mean of FIs computed for different directions. (**B**) Apparent photon-scaled mean-squared error for the estimated separation compared to the "ground truth" provided by triangulation from the estimated position of a fiducial bead. Shaded regions are bounded by the 25$^{th}$ and 75$^{th}$ percentiles, while solid lines demarcate the median. Colors are the same as in (A).

# Supplementary Materials for

## Quantum-inspired super-resolution of fluorescent point-like sources

Cheyenne S. Mitchell, Dhananjay Dhruva, Zachary P. Burke, David J. Durden, Armine I. Dingilian, Mikael P. Backlund
Corresponding author: mikaelb@illinois.edu

**The PDF file includes:**

## Materials and Methods

Theory and Computation

Simulated emitted fields and corresponding images due to isotropic point sources were generated using custom MATLAB routines at the level of classical vectorial diffraction theory. We take the emission to be quasi-monochromatic with wavelength $\lambda$ = 692 nm to match the center of the passband in our experiments. These calculations were based on previously published theory of dipole emitters embedded in and near various layered or homogeneous media (*1*–*3*). As described in Ref. (*3*), the electric field collected at the back focal plane of the microscope due to emission from a single dipolar source with transition electric dipole moment vector $\boldsymbol{\mu}$ (Fig. S1A) given by:

$$\boldsymbol{\mu} = \mu \begin{pmatrix} \cos\Phi \sin\Theta \\ \sin\Phi \sin\Theta \\ \cos\Theta \end{pmatrix} \tag{S1}$$

can be expressed in terms of a Green's tensor $\overleftrightarrow{\mathbf{G}}(\rho,\phi)$ according to:

$$\mathbf{E}_{\mathbf{BFP}}(\rho,\phi) = \overleftrightarrow{\mathbf{G}}(\rho,\phi)\cdot\boldsymbol{\mu} \tag{S2}$$

where $(\rho, \phi)$ denote the polar coordinates defined at the back focal plane (Fig. S1B). The elements of the Green's tensor for a source embedded above an air-glass interface were calculated from the equations given in Ref. (*1*). We assume collection with an index-matched oil immersion objective of NA = 1.45 to mimic our experiment. Any downstream linear optical processing (e.g. due to transmission through polarizing elements, phase elements, reflecting elements, and beam splitters) can be captured by a corresponding transformation on $\overleftrightarrow{\mathbf{G}}$. The field at the final image plane, $\mathbf{E}_{\mathbf{img}}(\rho',\phi')$, can then be expressed in terms of a transformed Green's tensor $\overleftrightarrow{\mathbf{G}}_{\mathbf{img}}$ (*3*):

$$\mathbf{E}_{\mathbf{img}}(\rho',\phi') = \overleftrightarrow{\mathbf{G}}_{\mathbf{img}}(\rho',\phi')\cdot\boldsymbol{\mu}\,. \tag{S3}$$

The corresponding image formed is given by:

$$I(\rho',\phi') = \mathbf{E}^{\dagger}_{\mathbf{img}}\mathbf{E}_{\mathbf{img}}\,. \tag{S4}$$

Equation (S4) can be rearranged to express the resulting image as an expansion in the elements of the orientational second-moment tensor $\boldsymbol{\mu}\boldsymbol{\mu}^{\dagger}$ (*3*). The image due to an isotropic point source can be obtained by integrating Eq. (S4) over all possible dipole orientations $(\Theta,\Phi)$ with equal weights. Since

$$\int_0^{\pi} \mathrm{d}\Theta \int_0^{2\pi} \mathrm{d}\Phi \sin\Theta\, \boldsymbol{\mu}(\Theta,\Phi)\boldsymbol{\mu}^{\dagger}(\Theta,\Phi) = \frac{4\pi}{3}\mathbf{I}_{\mathbf{3}}\,, \tag{S5}$$

where $\mathbf{I}_3$ is the 3x3 identity matrix, the image of an isotropic point source is equivalent to the incoherent sum of images due to *x*-, *y*-, and *z*-oriented dipoles.

The QFI associated with isotropic emission was computed by first associating the classical electric field generated at the back focal plane by an *x*-, *y*- and *z*-oriented dipole located at position (*x*, *y*) in the object plane with a corresponding normalized state vector representing the one-photon state generated by an oriented quantum emitter, as in Ref. (*4*). To facilitate numerical convergence, we transformed this state vector from the transverse momentum basis (equivalent to position on the Fourier plane) to the basis of Zernike polynomials truncated at the $45^{th}$ term. The density operator of the one-photon state generated by an isotropic emitter can then be obtained as an incoherent mixture of these state vectors. The density operator of the one-photon state generated by a pair of isotropic emitters at different positions can in turn be produced from the incoherent mixture of two such individual operators. With this final density operator in hand, numerical evaluation of the QFI becomes straightforward.

Sample Preparation

Samples of 40 nm fluorescent beads (Thermo Fisher Scientific F8789) were diluted 1:100 into an aqueous polyvinyl alcohol (PVA) solution (1% by weight). Microscope coverslips were cleaned using an UV-ozone etcher before aliquoting sample on top and spin coating for 1 min at 2000 rpm. Samples were then left to dry for a few minutes before being placed onto the microscope. The spin coating process resulted in a thin (~90 nm as measured by ellipsometry) layer of polymer, on top of which the sources sat fixed in place. That the beads sit on top of the film rather than within was determined by careful comparison of the experimental images to diffraction calculations that placed the emitter either above or below an air-glass interface. Some of the data depicted in Fig. S9 was recorded with a sample consisting of 70-nm nanodiamonds (Sigma-Aldrich 798169) spun in PVA rather than fluorescent beads.

Experimental Setup

The experimental setup is sketched in Fig. 2. Bead samples were placed on a PINano stage (Physik Instrumente) and pumped by a 640-nm laser (Coherent OBIS LX 1178790) passed through a quarter wave plate (Edmund Optics 39-038) to obtain approximately circularly polarized light. The intensity of the laser was ~30 $W/cm^2$ at the sample. We found that the polarization of the emitted light did not show much dependence on the polarization of the excitation light, perhaps due to coherent redistribution of energy within the bead via homo-FRET. The sample was illuminated in an epi-geometry, then fluorescence was collected using an NA 1.45/oil immersion objective (Olympus, UPLXAPO100X OIL OBJECTIVE, OFN26.5, NA1.45, WD0.13mm). Collected fluorescence was spectrally filtered through a dichroic mirror (IDEX FL-006889), a multi-stop notch filter (IDEX FL-009336), and a 692/40 bandpass filter (Edmund Optics 67038). The light exits the microscope body through a 200-mm focal length Nikon tube lens (Edmund Optics 58-520), then is passed through a second identical dichroic mirror (IDEX FL-006889) to compensate for residual birefringence (*5*). Several achromatic doublet lenses are used throughout the III to relay conjugate Fourier and image planes to desired locations. Referring to Fig. 2, these lenses are labeled $L_1$ (focal length = 300 mm, Edmund Optics 49-280-INK), $L_2$ (focal length = 100 mm, Edmund Optics 63-713-INK), and $L_3$ (focal length = 400 mm, Edmund Optics 49-281-INK). After passing through $L_1$ the fluorescence is directed through a vortex half wave plate (ThorLabs WPV10L-705) which converts azimuthally to vertically polarized light and radially to horizontally polarized light. A deformable mirror device (DMD, ThorLabs DMP40-P01) is placed at a conjugate Fourier plane in order to compensate low-order phase aberrations. The settings on the DMD were periodically reset (roughly every few days), but typical inputs are summarized in Table S1. After aberration correction, the light is filtered through a linear polarizer (ThorLabs LPVISE100-A) oriented to reject horizontally (initially radially) polarized light. After passing through $L_2$, the fluorescence enters the interferometer through a 50/50 beam splitter (ThorLabs BS013). In one interferometer arm the scene is inverted in the vertical direction by a vertically-oriented Dove prism (ThorLabs PS992M-A). In the other interferometer arm the scene is inverted in the horizontal direction using a horizontally oriented Dove prism. Both arms contain a delay stage composed of a right-angle prism (ThorLabs MRA25L-E02) and retroreflector (ThorLabs HRS1015-P01). In one arm the retroreflector is mounted on a piezoelectric translation stage (ThorLabs NFL5DP20) to enable dynamic control of the relative phase delay. The light is recombined at another 50/50 beam splitter, redirected with a knife-edge prism (ThorLabs MRAK25-G01), and focused by $L_3$ onto a water-cooled sCMOS camera (Hamamatsu ORCA-Fusion BT, C15440-20UP). Images were collected with the camera operating in ultra-quiet mode and binning 4x4 pixels, with an exposure time of 100 ms. The magnification in each of the four output images was carefully calibrated using a Ronchi ruling (400 lpmm, Edmund Optics), resulting in effective pixel sizes of 60.7, 60.3, 60.3, and 60.3 nm.

Experimental Protocol

Upon identification of a suitable target-fiducial bead pair, 50 calibration images were recorded in which the target was well separated from its inverted counterpart (Fig. S4A-B). The interferometer was then manually retuned by overlapping the image of the target bead with its inverted counterpart and then adjusting the delay stage to maximize the fringe visibility. A custom MATLAB routine was used to continuously record tens-of-thousands of images while the stage was scanned over a 200 nm x 200 nm area (Figs. S4C-D, S5). Table S2 summarizes additional experimental details including the average number of photons collected from the target per image pair for each of the experimental trials (on the order of $10^3$).

Data Analysis

Raw images were analyzed using a pipeline of bespoke MATLAB programs. First the relative positions of target and fiducial as they appear in each of four image copies was calibrated from the aforementioned set of 50 images in which the target image is well separated from its inverted counterpart (Fig. S4A-B). We refer to each of the four image copies as SL, SR, DL, and DR, where L/R refers to the left or right output port of the second beam splitter, and S/D refers to the output port of the first beam splitter (S for the interferometer arm with the "static" delay line, D for the one with the "dynamic" delay line). An estimate of the position of the fiducial as viewed in each of SL/SR/DL/DR was generated by least-squares fitting to a simplified model PSF. The same fitting was applied to each well-resolved image copy of the target bead. Ordinary direct images were fit to a Gaussian model of the form:

$$I_k^{(\mathrm{model})} = Ae^{-\left[(x_k-x_0)^2+(y_k-y_0)^2\right]/(2\sigma^2)} + b\,, \tag{S6}$$

where $I_k$, $x_k$, $y_k$ refer to the intensity, $x$ position, and $y$ position of the $k^{\mathrm{th}}$ pixel, and $A$, $x_0$, $y_0$, $\sigma$, and $b$ are the free parameters of the fit. Donut-PSF images were fit to the model:

$$I_k^{(\mathrm{model})} = Ae^{-\left[(x_k-x_0)^2+(y_k-y_0)^2\right]/(2\sigma^2)}\left[\frac{(x_k-x_0)^2+(y_k-y_0)^2}{2\sigma^2}\right]+b\,, \tag{S7}$$

with the same definitions as above. From these fits a set of relative position vectors connecting the target and fiducial in each of the four image copies was determined:

$$\mathbf{r}_{\mathbf{0rel}}^{\mathbf{S/D|L/R}} = \mathbf{r}_{\mathbf{0tar}}^{\mathbf{S/D|L/R}} - \mathbf{r}_{\mathbf{0fid}}^{\mathbf{S/D|L/R}}\,. \tag{S8}$$

Next, each image of the 2D scan was analyzed, where the images of the fiducials were still well-resolved but the images of the target overlapped significantly with their inverted counterparts (Fig. S4C-D). Estimates of the fiducial positions in each image copy, $\hat{\mathbf{r}}_{\mathbf{fid}}^{\mathbf{S/D|R/L}}$, were generated via least-squares fit to either Eq. (S6) or (S7). Using the relative positions determined in the calibration step, an estimate of the position of each image copy of the target could be triangulated:

$$\hat{\mathbf{r}}_{\mathbf{tar}}^{\mathbf{S/D|R/L}} = \hat{\mathbf{r}}_{\mathbf{fid}}^{\mathbf{S/D|R/L}} + \mathbf{r}_{\mathbf{0rel}}^{\mathbf{S/D|R/L}}\,, \tag{S9}$$

from which a set of separations between each target image and its inverted counterpart were computed:

$$\begin{aligned} \Delta x_L &= x_{\mathrm{tar}}^{\mathrm{SL}} - x_{\mathrm{tar}}^{\mathrm{DL}} \\ \Delta y_L &= y_{\mathrm{tar}}^{\mathrm{SL}} - y_{\mathrm{tar}}^{\mathrm{DL}} \\ \Delta x_R &= -x_{\mathrm{tar}}^{\mathrm{SR}} + x_{\mathrm{tar}}^{\mathrm{DR}} \\ \Delta y_R &= y_{\mathrm{tar}}^{\mathrm{SR}} - y_{\mathrm{tar}}^{\mathrm{DR}} \end{aligned}\,. \tag{S10}$$

The sign flip in the definition of $\Delta x_R$ accounts for the additional reflection incurred between the output ports of the second beam splitter. Results from the left and right images were subsequently averaged to give:

$$\begin{aligned} \Delta x &= \frac{1}{2}\left(\Delta x_L + \Delta x_R\right) \\ \Delta y &= \frac{1}{2}\left(\Delta y_L + \Delta y_R\right) \\ \Delta r &= \sqrt{\Delta x^2 + \Delta y^2} \end{aligned}\,. \tag{S11}$$

The result of the preceding analysis is a bank of tens-of-thousands of noisy, 19x19 images pairs of the target we denote $I_k^{(j,l)}\left(x_k, y_k; \Delta x_l, \Delta y_l\right)$, where $k$ refers to the pixel number, $j \in \{1, 2\}$ indicates the channel number (i.e. output port of the second beam splitter), and *l* runs from 1 to the total number of image sets in the bank.

To build image libraries from these data, such as the ones sampled in Figs. 3, S6, and S7, the noisy image banks were effectively low-pass filtered in several heuristically-tuned steps. First, to each individual background-corrected image we applied a Gaussian filter using the MATLAB function imgaussfilt(), with a specified width of 0.5 pixels. Next, in each of the 2x19x19 = 722 pixels comprising the image pairs of the target, we considered the fraction of total photons collected in that particular pixel as a function of $\left(\Delta x, \Delta y\right)$. This produced a 3D scatter plot with as many data points as there are image pairs recorded. The 3D scatter plot was fit to a quartic polynomial using the MATLAB fit() function with "fitType" set to "poly44". This fit effectively smooths the scatter plot and allows for interpolation between sampled $\left(\Delta x, \Delta y\right)$. The fit polynomial was then evaluated on a grid of finely sampled separations, ($\Delta x$, $\Delta y$), producing a look-up-table (LUT) of denoised image pairs we will denote $\mathcal{I}_k^{(1)}(x_k, y_k; \Delta x, \Delta y)$ and $\mathcal{I}_k^{(2)}(x_k, y_k; \Delta x, \Delta y)$, where index *k* still enumerates the pixel of the image. The coordinates ($\Delta x$, $\Delta y$) should be interpreted as the position vector separating the image of a single source from its inverted dual. The midpoint of this vector coincides with the optical axis. If the PSF were perfectly circularly symmetric then the denoised image at a given ($\Delta x$, $\Delta y$) should be entirely equivalent (within a constant factor) to the noiseless image of two equally bright, mutually incoherent point sources separated by ($\Delta x$, $\Delta y$), with their centroid coaligned to the optical axis. Put a slightly different way, for a circularly symmetric PSF we expect $\mathcal{I}_k^{(j)}(x_k, y_k; \Delta x, \Delta y) = \mathcal{I}_k^{(j)}(x_k, y_k; -\Delta x, -\Delta y)$. While theory certainly predicts a circularly symmetric PSF, we did not find this to be the case in practice (as can be seen by careful inspection of Fig. S4A-B), and so in fact $\mathcal{I}_k^{(j)}(x_k, y_k; \Delta x, \Delta y) \neq \mathcal{I}_k^{(j)}(x_k, y_k; -\Delta x, -\Delta y)$. A symmetrized LUT was defined:

$$\overline{\mathcal{I}_k^{(j)}}(x_k, y_k; \Delta x, \Delta y) = \frac{1}{2}\left[\mathcal{I}_k^{(j)}(x_k, y_k; \Delta x, \Delta y) + \mathcal{I}_k^{(j)}(x_k, y_k; -\Delta x, -\Delta y)\right]. \qquad \text{(S12)}$$

The images comprising the symmetrized LUT were then again passed through a Gaussian blur using imgaussfilt() with a width of 0.5 pixels. The images in Figs. 3, S6, and S7 are samples of these $\overline{\mathcal{I}_k^{(j)}}$. Recovered FIs (i.e. solid lines in Figs. 4A and S10) were calculated directly from these $\overline{\mathcal{I}_k^{(j)}}$.

Finally, we detail the method by which we generated the MSE data plotted in Figs. 4B and S11. As described above, in our scanning experiments we recorded large banks of noisy images, denoted $I_k^{(j,l)}\left(x_k, y_k; \Delta x_l, \Delta y_l\right)$, of single point-like sources displaced from the optical axis by an estimated position $\left(\Delta x_l, \Delta y_l\right)$. To simulate noisy images of two point sources, for each *l* we searched the image bank for the noisy image with estimated displacement closest to $\left(-\Delta x_l, -\Delta y_l\right)$, i.e. we determined $l_0$ for which:

$$l_0 = \arg\min_{l'} \left[ \sqrt{\left(-\Delta x_l - \Delta x_{l'}\right)^2 + \left(-\Delta y_l - \Delta y_{l'}\right)^2} \right]. \tag{S13}$$

In most cases we were able to couple each $\left(\Delta x_l, \Delta y_l\right)$ with a corresponding $\left(\Delta x_{l_0}, \Delta y_{l_0}\right)$ such that $\sqrt{\left(-\Delta x_l - \Delta x_{l_0}\right)^2 + \left(-\Delta y_l - \Delta y_{l_0}\right)^2}$ was at most a few nanometers. The fact that no match will be perfect emulates something like a misalignment of the optical axis to the exact centroid of the source pair (*6*). Defining

$$\begin{aligned} \Delta x_l &= \frac{1}{2}\left(\Delta x_l - \Delta x_{l_0}\right) \\ \Delta y_l &= \frac{1}{2}\left(\Delta y_l - \Delta y_{l_0}\right) \quad , \\ \Delta r_l &= \sqrt{\left(\Delta x_l\right)^2 + \left(\Delta y_l\right)^2} \end{aligned} \tag{S14}$$

noisy images of simulated source pairs were generated according to:

$$\begin{aligned} \overline{I_k^{(j,l)}}\left(x_k, y_k; \Delta x_l, \Delta y_l\right) = \quad & I_k^{(j,l)}\left(x_k, y_k; \Delta x_l, \Delta y_l\right) \left[ \frac{\sum_k I_k^{(j,l)}\left(x_k, y_k; \Delta x_l, \Delta y_l\right) + \sum_k I_k^{(j,l_0)}\left(x_k, y_k; \Delta x_{l_0}, \Delta y_{l_0}\right)}{2\sum_k I_k^{(j,l)}\left(x_k, y_k; \Delta x_l, \Delta y_l\right)} \right] \\ & + I_k^{(j,l_0)}\left(x_k, y_k; \Delta x_{l_0}, \Delta y_{l_0}\right) \left[ \frac{\sum_k I_k^{(j,l)}\left(x_k, y_k; \Delta x_l, \Delta y_l\right) + \sum_k I_k^{(j,l_0)}\left(x_k, y_k; \Delta x_{l_0}, \Delta y_{l_0}\right)}{2\sum_k I_k^{(j,l_0)}\left(x_k, y_k; \Delta x_{l_0}, \Delta y_{l_0}\right)} \right] \end{aligned}$$

(S15)

where the factors in square brackets enforce equal brightness of the two sources. In the computation of MSE described below, $\Delta r_l$ is taken as the "ground-truth" separation of the source pair, though it itself is only an estimate and contains error terms, for example due to the stochastic localization of the fiducial and/or imperfect registration of the four image copies.

Next we generated another estimate of separation we denote $\Delta r_l$ via one of two basic ways. For the direct imaging data depicted in Figs. 4B (blue) and S11 (blue, solid and dashed) we computed the least-squares fit of each $\overline{I_k^{(j,l)}}$ to a model comprised of the sum of two Gaussian functions of the form given in Eq. (S6). $\Delta r_l$ was then equated with the distance between the centers of these two fit Gaussians. As summarized in Table S2, images for each microscope configuration were recorded over the course of two independent "trials" conducted on different days. In the second method of separation estimation used to produce the red lines in Fig. 4B as well as the red, blue dotted, and blue dashed-dotted lines in Fig. S11, the images from one trial were used to generate a LUT, $\overline{\mathcal{I}_k^{(j)}}(x_k, y_k; \Delta x, \Delta y)$, as described above. Then the data from the other, independent trial were used to produce $\overline{I_k^{(j,l)}}\left(x_k, y_k; \Delta x_l, \Delta y_l\right)$. A new estimate of separation was generated by searching the LUT for the image pair minimizing the mean-squared error, i.e.:

$$\mathbf{\Delta r}_l = \arg\min_{\mathbf{\Delta r}} \left\{ \sum_{j,k} \left[ \overline{\mathcal{I}_k^{(j)}}(x_k, y_k; \Delta x, \Delta y) - \overline{I_k^{(j,l)}}\left(x_k, y_k; \Delta x_l, \Delta y_l\right) \right]^2 \right\}. \tag{S16}$$

The MSE was computed by binning $\Delta r_l$ into 5-nm segments, then averaging the squared error $\left[\Delta r_l - \Delta r_l\right]^2$ over all $l$ in the bin. Again we note that since the "ground truth" in fact carries error terms, the MSE as defined here includes a component from this term alone. In effect this will raise the floor of the plotted MSEs.

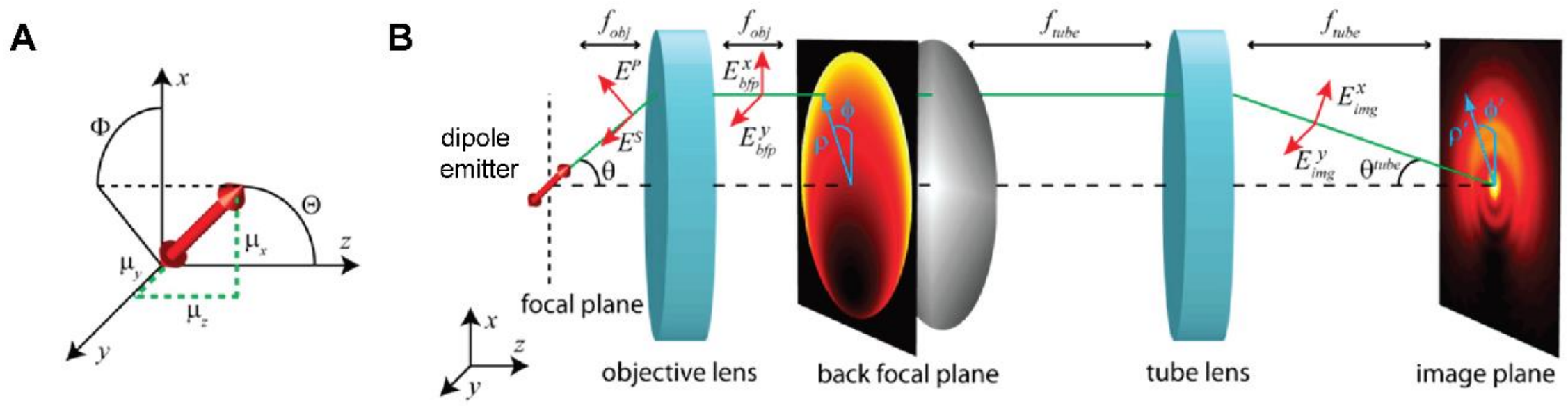

**Fig. S1.**

Coordinate definitions for discussion on the effect of dipolar emission. (**A**) Dipole vector $\boldsymbol{\mu} = [\mu_x, \mu_y, \mu_z]^\dagger = \mu[\cos\Phi\sin\Theta, \sin\Phi\sin\Theta, \cos\Theta]^\dagger$ defined in terms of the dipole's azimuthal, Φ, and polar, Θ, orientations. (**B**) Coordinate definitions relative to imaging system. Most importantly, *z* is taken to be parallel to the optical axis. Adapted from Fig. 2 of Ref. (*3*).

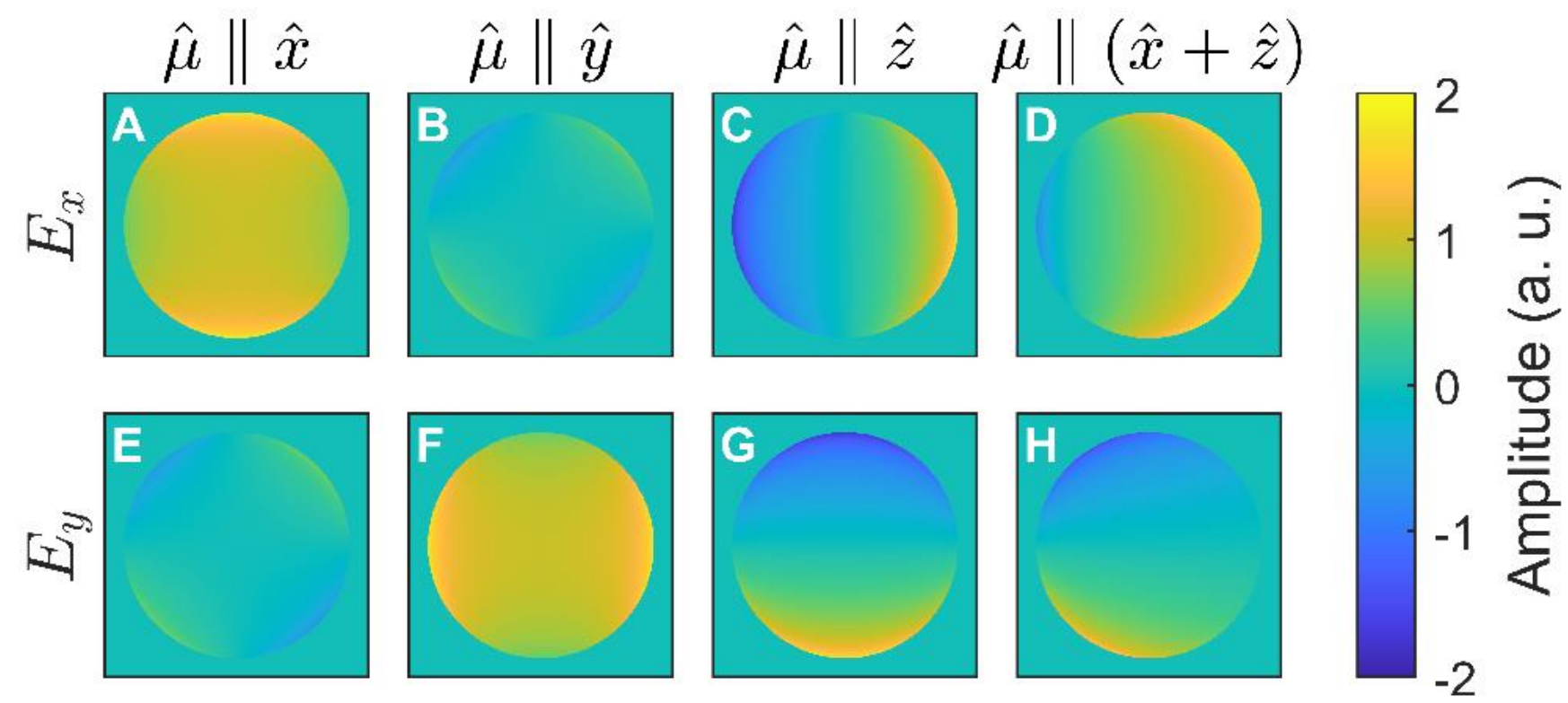

**Fig. S2.**

Simulated electric field distributions at the back focal plane of the microscope due to collected emission from a dipole oriented parallel to the *x* axis (**A, E**), parallel to the *y* axis (**B, F**), parallel to the *z* axis (**C, G**), and oriented in the *xz* plane at an angle 45° from the *z* axis (**D, H**). The fields depicted in (A-B, E-F) are symmetric with respect to inversion. Those in (C, G) are anti-symmetric with respect to inversion, whereas those in (D, H) are asymmetric.

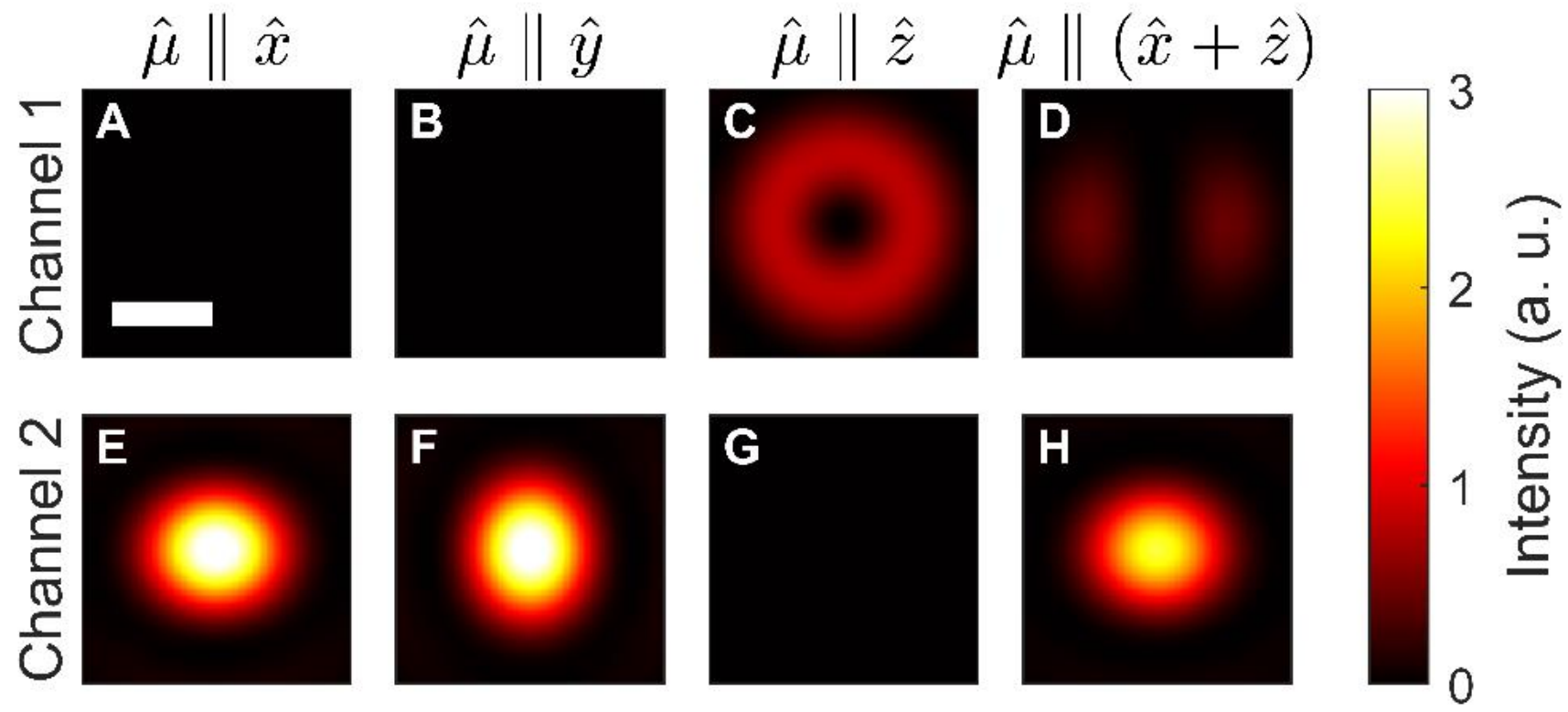

**Fig. S3.**

Simulated, high-resolution images of dipole emitters oriented parallel to the *x* axis (**A, E**), parallel to the *y* axis (**B, F**), parallel to the *z* axis (**C, G**), and oriented in the *xz* plane at an angle 45° from the *z* axis (**D, H**), as viewed at the output ports of an unpolarized image inversion interferometer. Due to symmetry considerations (Fig. S2), the output in Channel 1 is nulled in (A-B, E-F). The output in Channel 2 is nulled in (C, G). Neither Channel is nulled in (D, H). Scale bar: 250 nm.

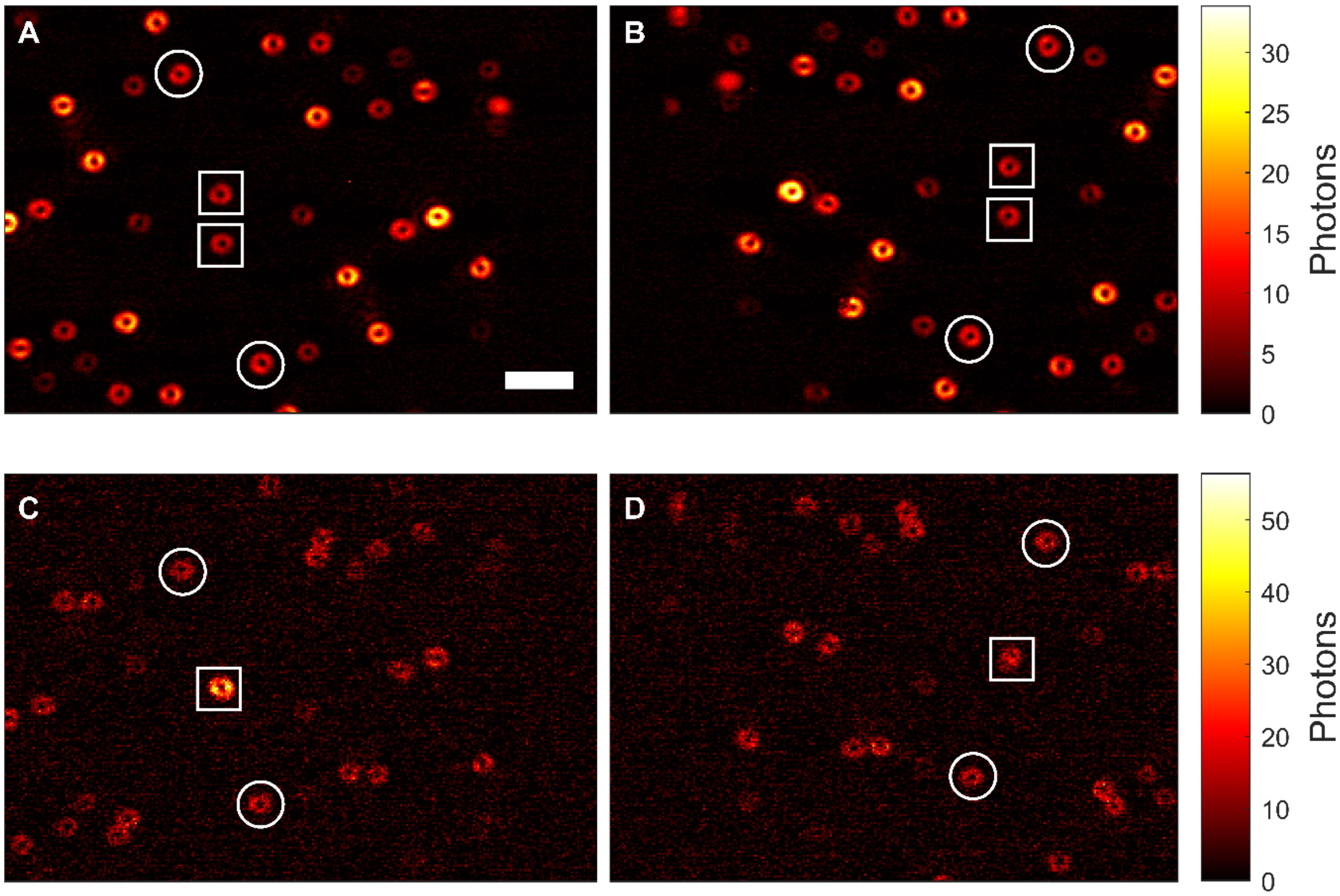

**Fig. S4.**

Example experimental images of a field-of-view containing several fluorescent beads with the microscope configured for vortex-y-polarized image inversion interferometry. (**A-B**) Average of 50 100-ms dark-count-corrected image frames from a fiducial calibration. Two image copies appear on the left side of the camera chip (A), and two appear on the right (B). Images copies within each panel are spatially inverted relative to one another. Images on the left and right are related by horizontal inversion due to an additional reflection at the output of the second beam splitter. White circles highlight the four image copies of the fluorescent bead that will be treated as the fiducial. White squares mark the four image copies of the fluorescent bead that will be treated as the target. (**C-D**) A single 100-ms dark-count-corrected image frame recorded during a scan of the same field-of-view. Copies of the PSF of the target (regions in white squares) now overlap heavily with one another, producing constructive interference in Channel 1 (C) and destructive interference in Channel 2 (D). Scale bar: 2 μm.

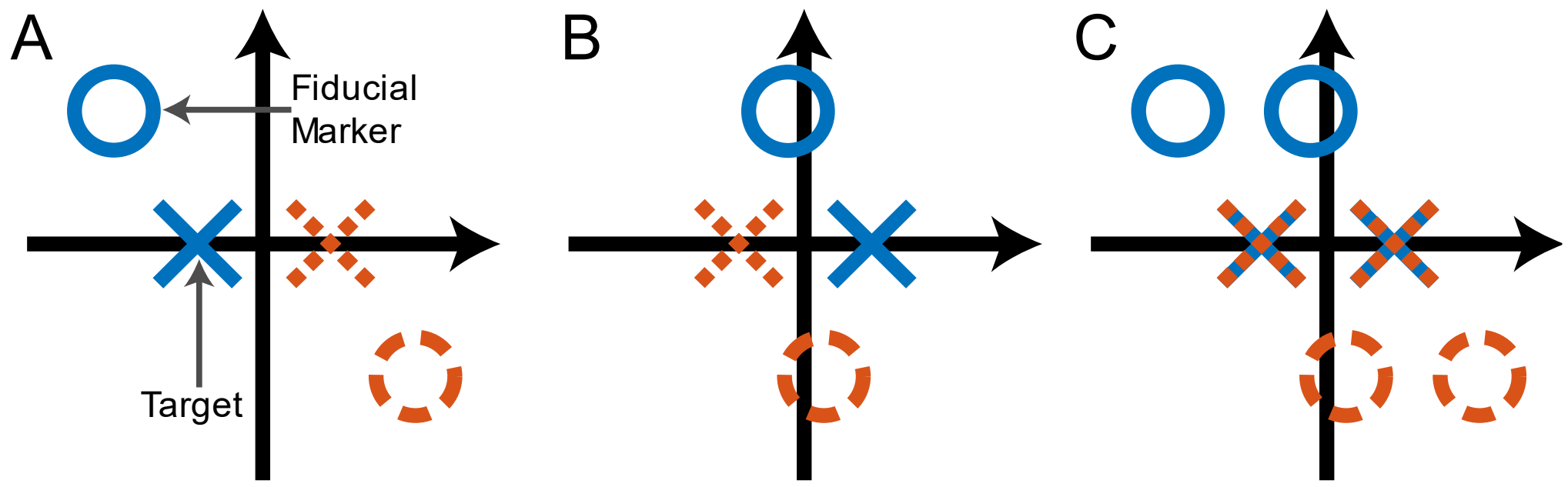

**Fig. S5.**
A schematic of the analog simulation of two mutually incoherent emitters, with “o” marking the fiducial and “x” the target. Black arrows intersect at the optical axis. Solid blue lines indicate a single image copy of the fiducial/target, whereas broken red lines mark its inverted dual. (**A**) Images are recorded with the target located some small displacement away from its inverted dual. (**B**) Some time later the stage is moved such that the target now appears at the opposite position as it did in (A). (**C**) The images from (A) and (B) are then combined in post-processing.

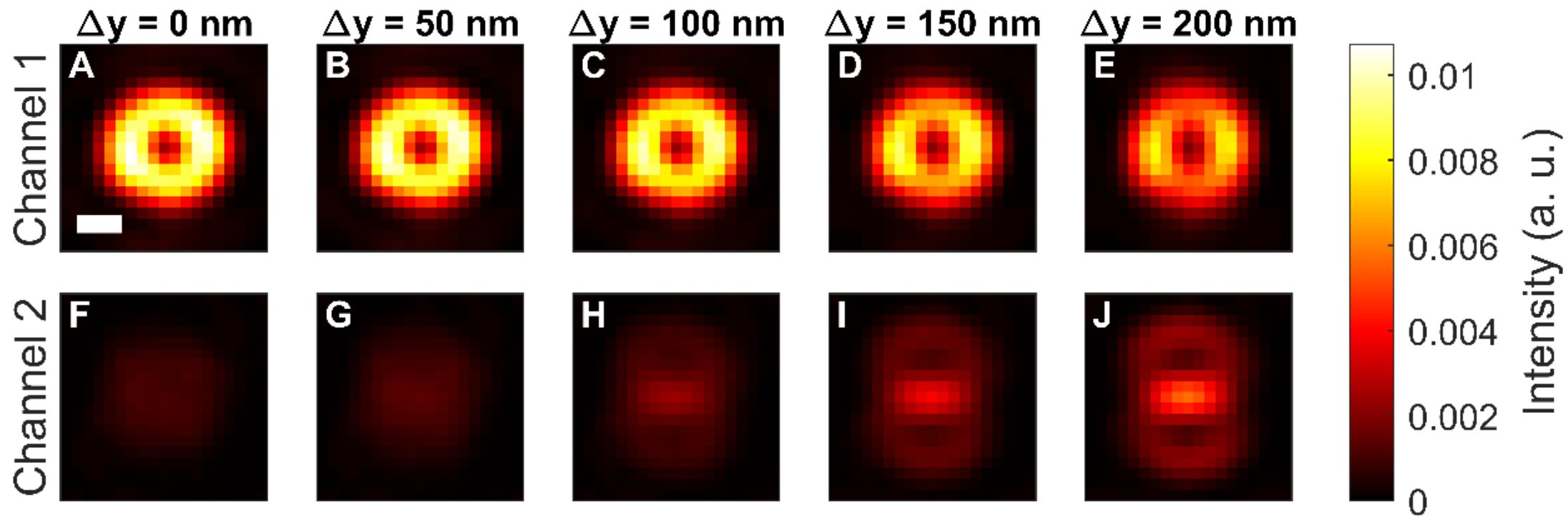

**Fig. S6.**

Processed images sampled from the same LUT as is depicted in Fig. 3A-E, but with the source separation along *y* rather than *x*. (**A-E**) Images output in Channel 1 of the vortex-y-polarized image inversion interferometer at separations (A) $\Delta r = \Delta y = 0$ nm, (B) $\Delta y = 50$ nm, (C) $\Delta y = 100$ nm, (D) $\Delta y = 150$ nm, and (E) $\Delta y = 200$ nm. (**F-J**) Images output in Channel 2 of the vortex-y-polarized image inversion interferometer at separations (F) $\Delta y = 0$ nm, (G) $\Delta y = 50$ nm, (H) $\Delta y = 100$ nm, (I) $\Delta y = 150$ nm, and (J) $\Delta y = 200$ nm. The "bowtie" is now oriented vertically, along the direction of separation. Scale bar: 250 nm.

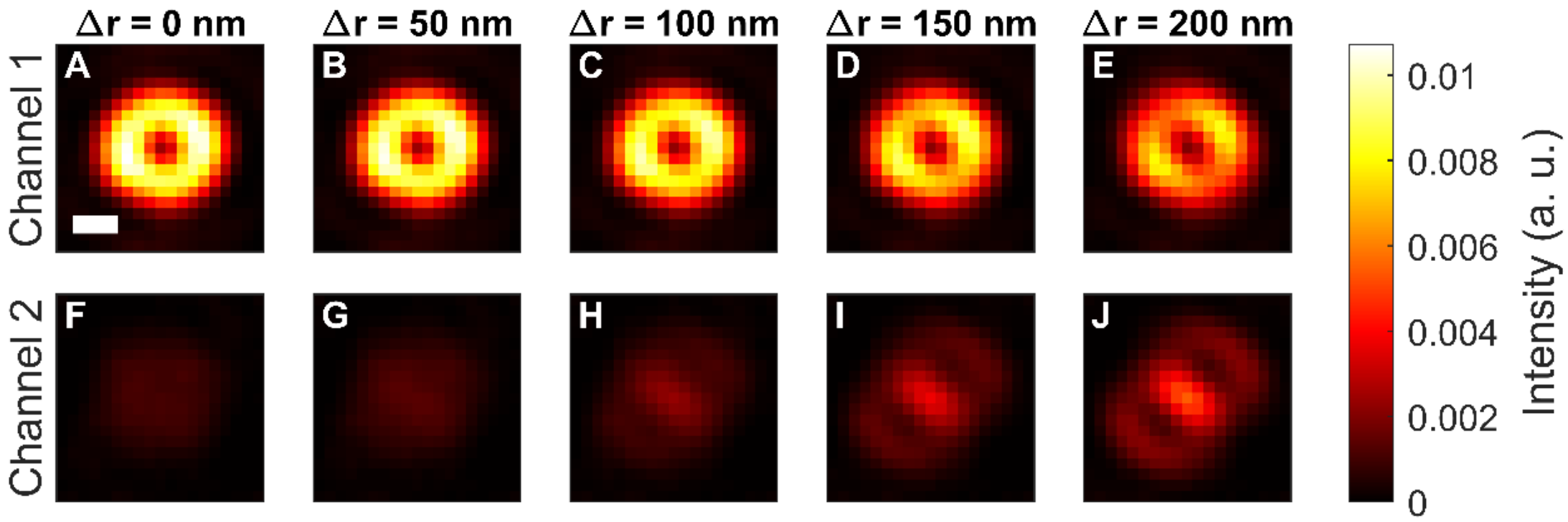

**Fig. S7.**

Processed images sampled from the same LUT as is depicted in Figs. 3A-E and S6, but with the source separation oriented along the image diagonal. (**A-E**) Images output in Channel 1 of the vortex-y-polarized image inversion interferometer at separations (A) Δ*r* = 0 nm, (B) Δ*r* = 50 nm, (C) Δ*r* = 100 nm, (D) Δ*r* = 150 nm, and (E) Δ*r* = 200 nm. (**F-J**) Images output in Channel 2 of the vortex-y-polarized image inversion interferometer at separations (F) Δ*r* = 0 nm, (G) Δ*r* = 50 nm, (H) Δ*r* = 100 nm, (I) Δ*r* = 150 nm, and (J) Δ*r* = 200 nm. Scale bar: 250 nm.

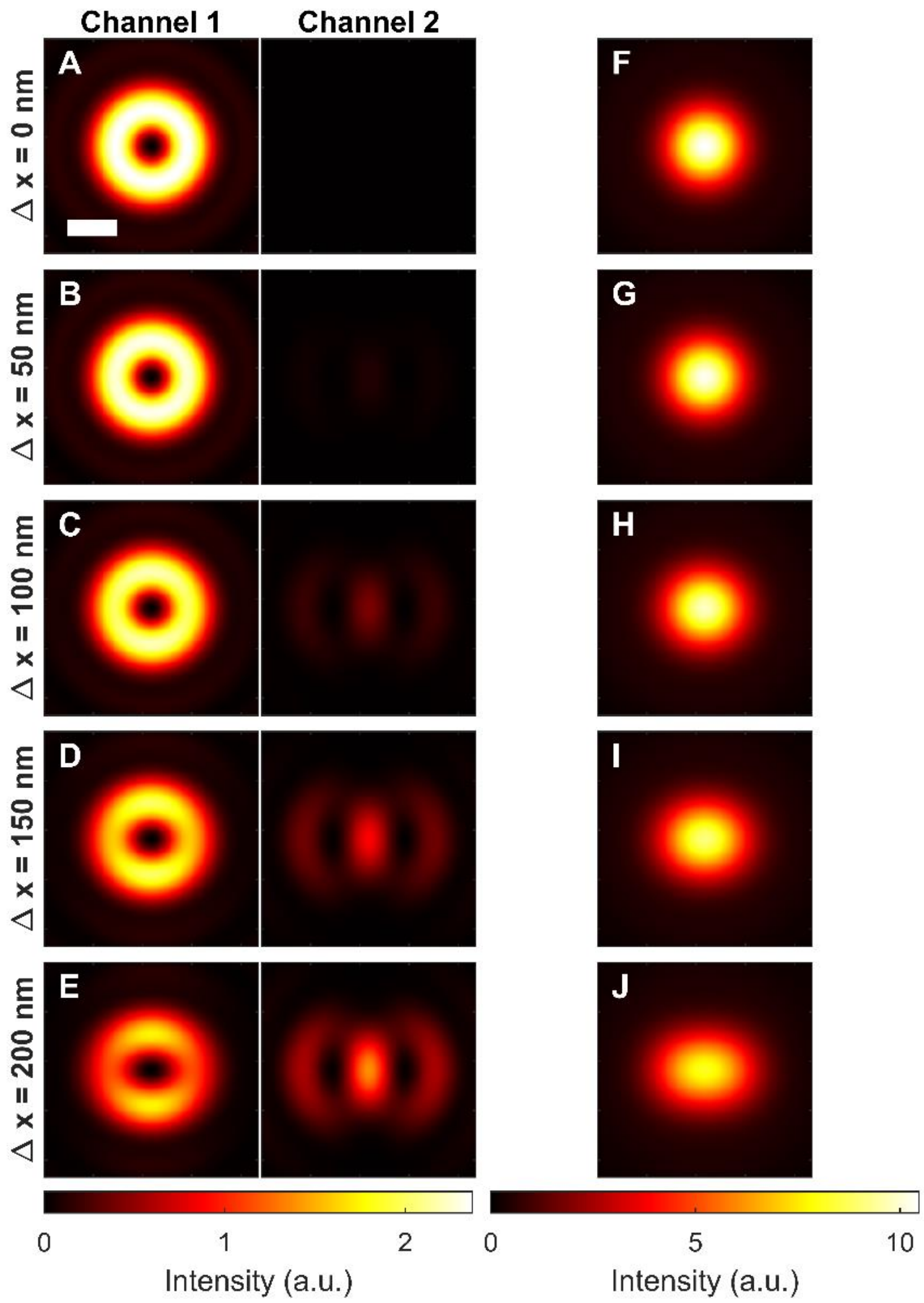

**Fig. S8.**

Simulated, high-resolution images of a pair of isotropic emitters at small separations. (**A-E**) Simulated images output in Channels 1 (left) and 2 (right) of the vortex-y-polarized III at separations (A) $\Delta r = \Delta x = 0$ nm, (B) $\Delta x = 50$ nm, (C) $\Delta x = 100$ nm, (D) $\Delta x = 150$ nm, and (E) $\Delta x = 200$ nm. These images can be compared directly to the experimentally-derived LUT images depicted in Fig. 3A-E. (**F-J**) Simulated images derived from ordinary direct imaging at separations (F) $\Delta r = \Delta x = 0$ nm, (G) $\Delta x = 50$ nm, (H) $\Delta x = 100$ nm, (I) $\Delta x = 150$ nm, and (J) $\Delta x = 200$ nm. These images can be compared directly to the experimentally-derived LUT images depicted in Fig. 3F-J. Scale bar: 250 nm.

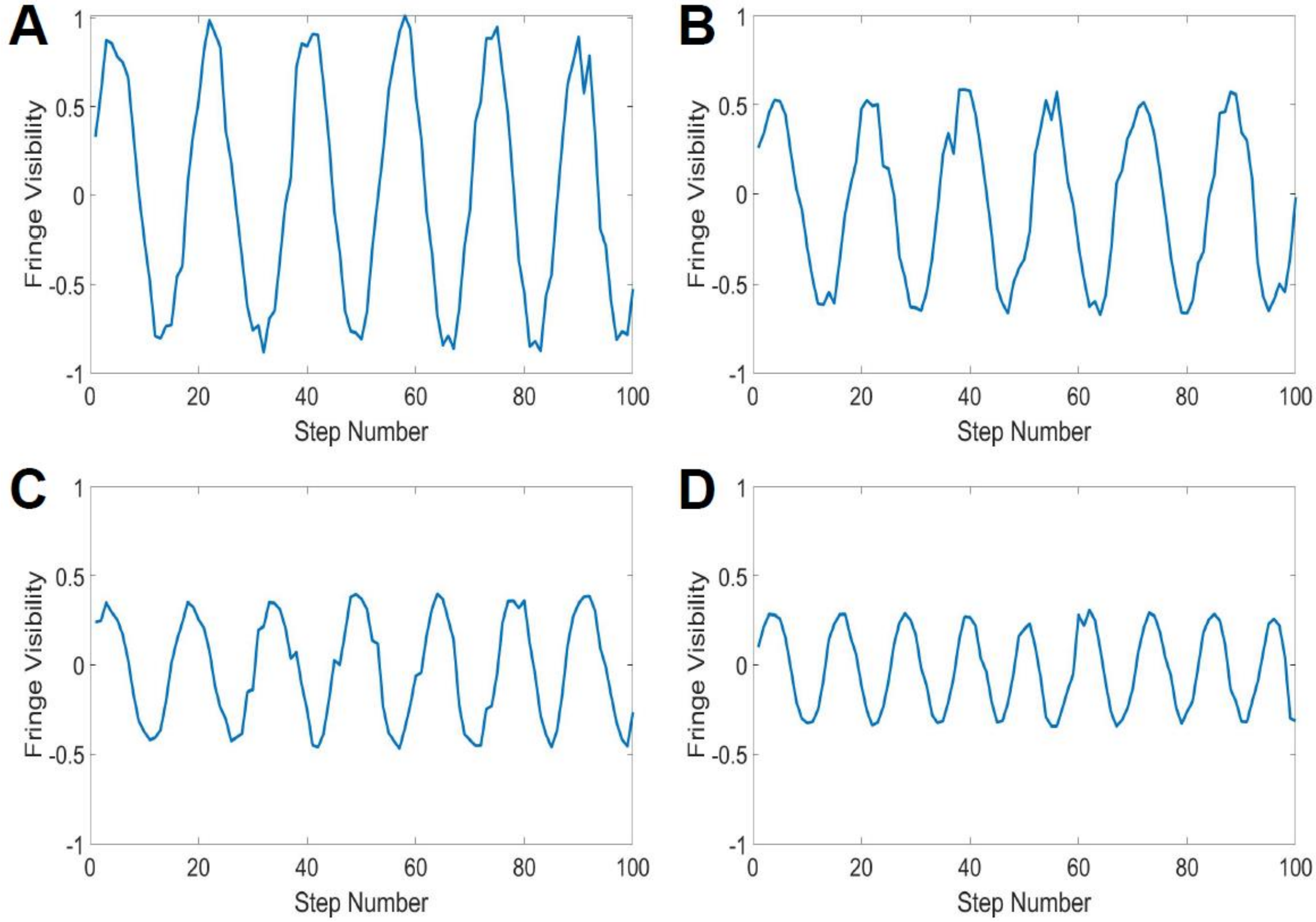

**Fig. S9.**

Experimentally-realized fringe visibility as a function of the relative delay between interferometer arms for the following four configurations: vortex-y-polarized III (**A**), linearly polarized III (**B**), unpolarized III with aberration corrections (**C**), and unpolarized III with no aberration corrections (**D**). The max visibility degrades monotonically from A to D.

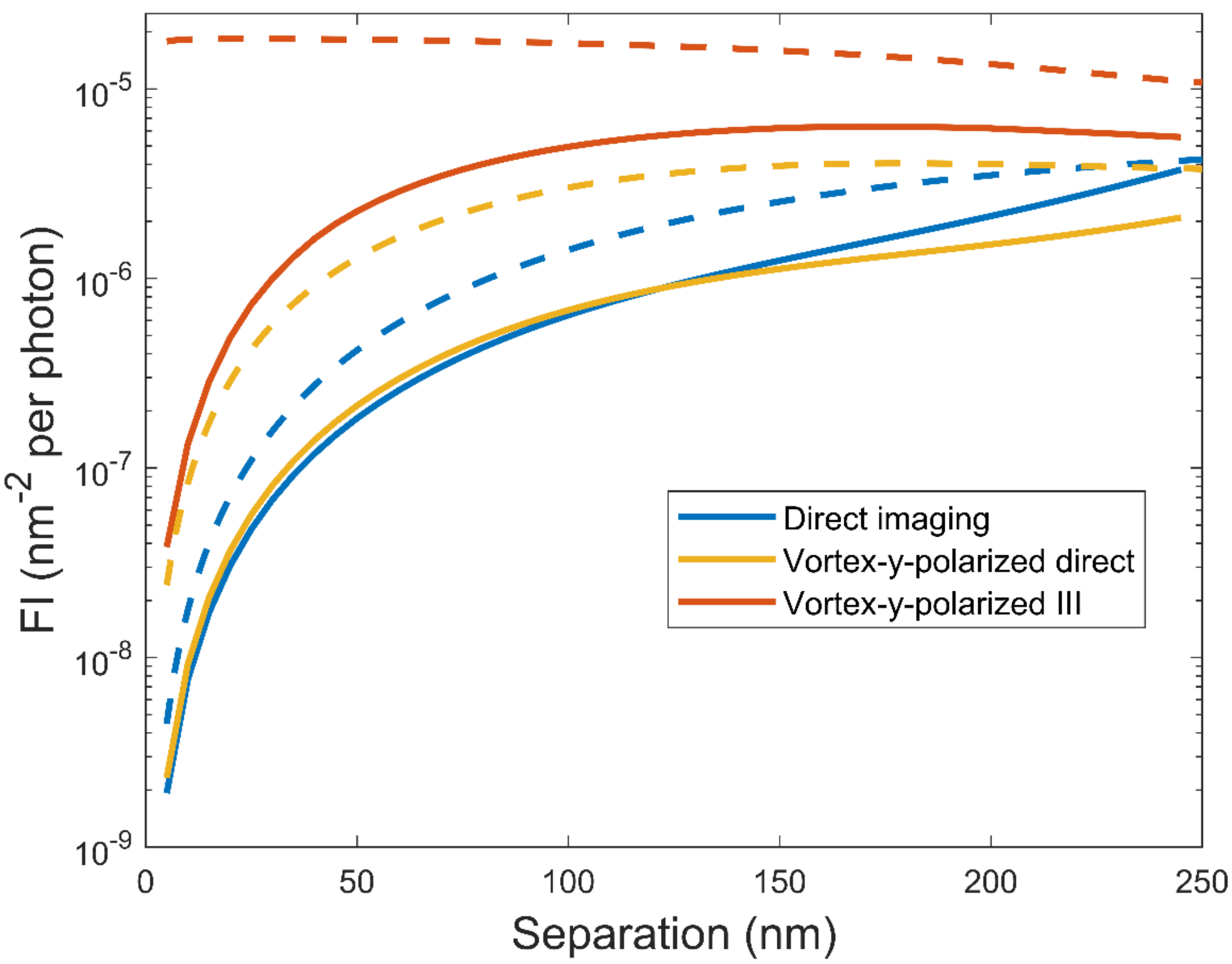

**Fig. S10.**

The realized Fisher information, as derived from processed experimental images of the vortex-y-polarized image inversion interferometer (red), ordinary direct imaging (blue), and direct imaging with the donut PSF (gold). Dashed lines indicate predictions from theory. The red and blue data appear already in Fig. 4A but are included again here for context.

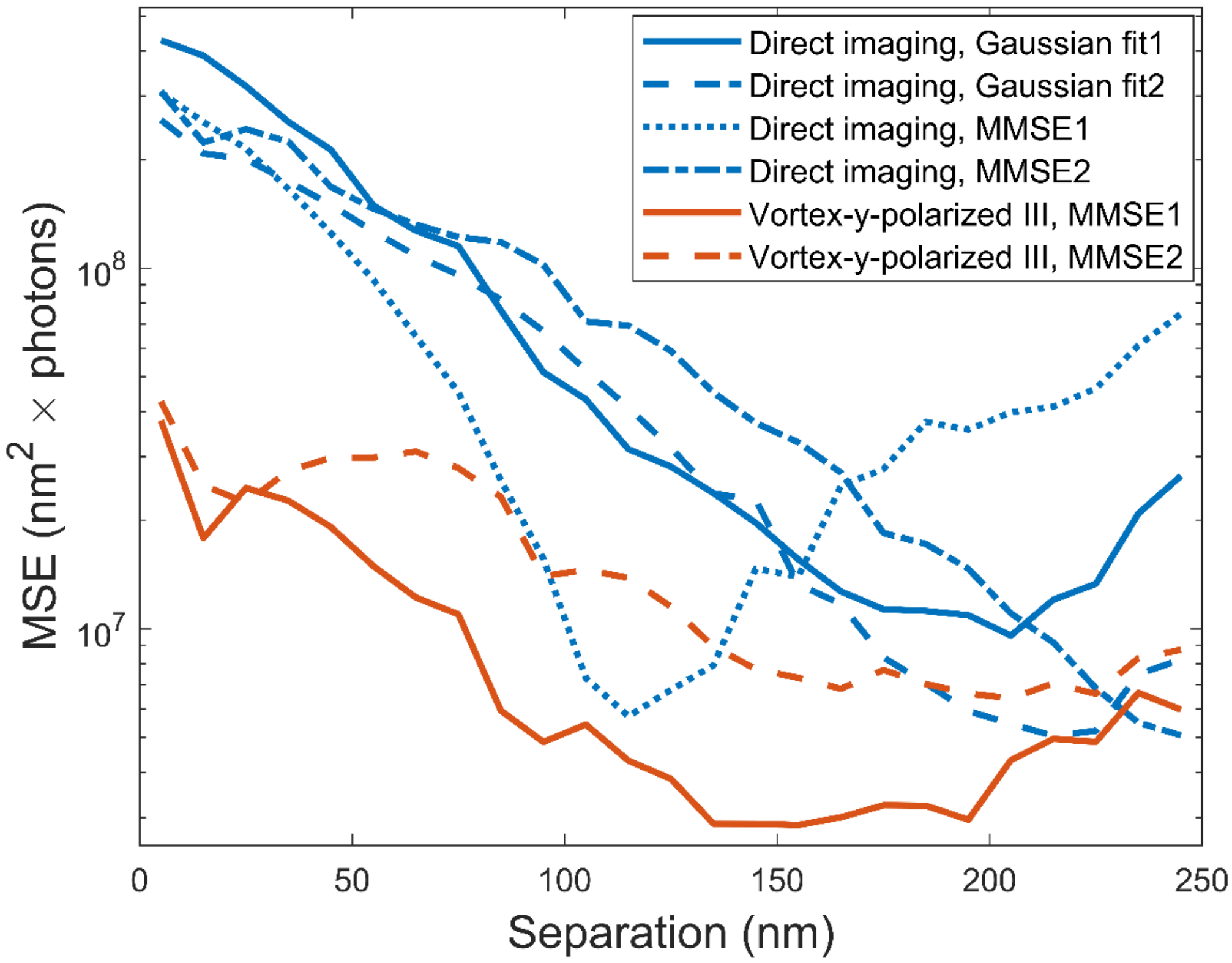

**Fig. S11.**

Experimental mean-squared error vs. separation, enumerated several different ways as detailed in the Materials and Methods section. A key observation is that at small separations, the blue lines are consistently about an order-of-magnitude higher than the red lines.

**Table S1.**

A typical set of inputs for the deformable mirror (ThorLabs DMP40-P01) implemented to correct for common-path phase aberrations, presumably resulting from imperfections in the optical elements themselves.

| **Zernike Polynomial** | **Relative Voltage** |
|---|---|
| Z4 | 0.03 |
| Z5 | 0.00 |
| Z6 | 0.01 |
| Z7 | -0.03 |
| Z8 | 0.05 |
| Z9 | 0.12 |
| Z10 | 0.01 |
| Z11 | 0.00 |
| Z12 | -0.01 |
| Z13 | 0.07 |
| Z14 | 0.01 |
| Z15 | 0.00 |

**Table S2.**

Summary of experimental configurations and trials. Each trial was recorded on a different day. An "image set" refers to a single 100-ms frame containing all four reflected and inverted copies of the same scene (i.e., SL/DL/SR/DR).

| Experimental condition | Trial | Number of target beads recorded | Number of image sets | Mean target photons per image set |
|---|---|---|---|---|
| Vortex-y-pol. III | 1 | 6 | 25056 | 4530 |
| Vortex-y-pol. III | 2 | 5 | 15789 | 2778 |
| Unpolarized direct | 1 | 4 | 15511 | 4143 |
| Unpolarized direct | 2 | 4 | 16481 | 1459 |
| Vortex-y-pol. direct | 1 | 6 | 24401 | 7205 |
| Vortex-y-pol. direct | 2 | 5 | 19784 | 5701 |